\documentclass[aip,10pt,twocolumn]{revtex4-2}

\usepackage{amsmath}

\usepackage[pdftex]{graphicx}

\begin{document}

\author{Paul M. Donaldson$^{1,*}$, Greg M. Greetham$^1$, Chris T. Middleton$^2$, Brad M. Luther$^3$, Martin T. Zanni$^4$, Peter Hamm$^{5,*}$, Amber T. Krummel$^{3,*}$\\\vspace{0.3cm}
$^1$Central Laser Facility, Research Complex at Harwell, STFC Rutherford Appleton Laboratory, Harwell Science and Innovation Campus, Didcot, OX11 0QX, UK\\
$^2$PhaseTech Spectroscopy, Inc., 4916 East Broadway, Suite 125, Madison, WI, USA\\
$^3$Colorado State University, Department of Chemistry, 200 W. Lake Street, Fort Collins, CO, USA\\
$^4$University of Wisconsin, Department of Chemistry, Room 8361, 1101 University Ave., Madison, WI, USA\\
$^5$University of Zurich, Department of Chemistry, Winterthurerstrasse 190, CH-8057, Zurich, Switzerland\\
$^3$Colorado State University, Department of Chemistry, 200 W. Lake Street, Fort Collins, CO, USA\\
\vspace{0.3cm}
$^*$corresponding authors: paul.donaldson@stfc.ac.uk, peter.hamm@chem.uzh.ch, amber.krummel@colostate.edu
}

\title {Breaking Barriers in Ultrafast Spectroscopy and Imaging Using 100 kHz Amplified Yb-Laser Systems}

\begin{abstract}


\noindent\textbf{Conspectus:} Ultrafast spectroscopy and imaging have become tools utilized by a broad range of scientists involved in materials, energy, biological, and chemical sciences.  Commercialization of ultrafast spectrometers including transient absorption spectrometers, vibrational sum frequency generation spectrometers, and even multidimensional spectrometers have put these advanced spectroscopy measurements into the hands of practitioners originally outside the field of ultrafast spectroscopy.  There is now a technology shift occurring in ultrafast spectroscopy, made possible by new Yb-based lasers, that is opening exciting new experiments in the chemical and physical sciences.  Amplified Yb-based lasers are not only more compact and efficient than their predecessors, but most importantly operate at many times the repetition rate with improved noise characteristics in comparison to the previous generation of Ti:Sapphire amplifier technologies. Taken together, these attributes are enabling new experiments, generating improvements to long-standing techniques, and affording the transformation of spectroscopies to microscopies. This account aims to show that the shift to 100~kHz lasers is a transformative step in nonlinear spectroscopy and imaging, much like the dramatic expansion that occurred with the commercialization of Ti:Sapphire laser systems in the 1990s. The impact of this technology will be felt across a great swath of scientific communities. We first describe the technology landscape of amplified Yb-based laser systems used in conjunction with 100~kHz spectrometers operating with shot-to-shot pulse shaping and detection. We also identify the range of different parametric conversion and supercontinuum techniques which now provide a path to making pulses of light optimal for ultrafast spectroscopy. Second we describe specific instances from our laboratories of how the amplified Yb-based light sources and spectrometers are transformative. For multiple probe time resolved infrared and transient 2D IR spectroscopy, the gain in temporal resolution and signal-to-noise enables dynamical spectroscopy measurements from femtoseconds to seconds. These gains widen the applicability of time resolved infrared techniques across a range of topics in photochemistry, photocatalysis and photobiology, as well as lowering the technical barriers to implementation in a laboratory. For 2D Visible spectroscopy and microscopy with white light, as well as 2D IR imaging, the high repetition rates of these new Yb-based light sources allows for spatially mapping 2D spectra while maintaining high signal-to-noise in the data.  To illustrate the gains, we provide examples of imaging applications in the study of photovoltaic materials and spectroelectrochemistry.
\end{abstract}

\maketitle

\section*{Key References}

\begin{itemize}
\item Luther, B.M., Tracy, K.M., Gerrity, M., Brown, S., and Krummel, A.T. 2D~IR Spectroscopy
at 100 kHz Utilizing a Mid-IR OPCPA Laser Source, Opt. Express, 2016, 24, 4117-4127.\cite{Luther2016} \textit{A description of a 100 kHz OPCPA laser system capable of producing tunable pulses generated in the 3-6 micron region and used to perform 2D~IR spectroscopy.}

\item Greetham, G. M., Donaldson, P.M., Nation, C., Sazanovich, I.V., Clark, I.P., Shaw, D.J., Parker, A.W., and Towrie, M., A 100 kHz Time-Resolved Multiple-Probe Femtosecond to Second Infrared Absorption Spectrometer, Appl. Spectrosc., 2016, 70, 645-653.\cite{Greetham2016} \textit{A 100 kHz dual Yb laser amplifier system, enabling transient IR and visible spectroscopy experiments from femtosecond to second timescales. The system is a UK national facility for scientific and industrial user access with broad application.}

\item Jones, A.C., Kearns, N.M., Bohlmann Kunz, M., Flach, J.T., and Zanni, M.T., Multidimensional spectroscopy on the microscale: Development of a multimodal imaging system incorporating 2D White-Light spectroscopy, broadband transient absorption, and Atomic Force Microscopy, J. Phys. Chem.y A, 2019, 123, 10824-10836.\cite{Jones2019b} \textit{Described the experimental apparatus for a 2D White-Light and broadband TA microscope with AFM topology correlation.}

\item Hamm, P., Transient 2D~IR Spectroscopy from Micro- to Milliseconds, J. Chem. Phys., 2021, 154, 104201.\cite{Hamm2021} \textit{A 100~kHz Yb laser is used to collect a sequence of many transient 2D~IR spectra separated by 10~$\mu$s each resolving the photocycle of bacterio rhodopsin.}

\end{itemize}

\section*{Motivation}

Many processes in physics, chemistry, and biology happen on ultrafast timescales, and understanding them requires that we are able to observe them. This is the realm of  time-resolved femtosecond spectroscopy. In these experiments, a physical or chemical process is initiated in a sample with an intense pulse of pump light, and then followed spectroscopically in time with probe light. This  family of techniques has provided vital data on a huge variety of light-induced chemical processes.\cite{zewail2000} With the invention of mode locking in dye lasers,\cite{Ippen1974} light pulses as short as 100~fs could be achieved, which were compressed down to 6~fs only a few years later.\cite{Brito-Cruz1987} Since then, a time-resolution sufficient for the ``speed limit'' of atomic motion is available.

The first technological shift in time-resolved femtosecond spectroscopy came about  with the invention of Ti:Sapphire (Ti:Sa) lasers as a source of femtosecond light pulses,\cite{Spence1991} combined with the Nobel prize winning development of Chirped Pulse Amplification (CPA) in the mid 1980’s.\cite{Strickland1985}  The reliability of femtosecond light sources improved significantly with Ti:Sa lasers, since the complete systems are based on solid-state lasers. In particular, laser diode pumps could provide pump energy directly into the desired transition of the laser active medium, improving electrical efficiency and decreasing thermal load on the gain material, which otherwise limits the average power of the laser.   As diode pumps improved, workhorse frequency-doubled Nd:YAG lasers saw improved performance as well.
Nonetheless, since their introduction about 30 years ago, Ti:Sa laser technology barely changed.

The next technological shift started about 10 years ago. Newly developed
amplified Yb-based laser systems have paved the way for significant improvements in time-resolved spectroscopies, and for new experiments in spectroscopy and imaging to become practical. There are several characteristics of these laser systems that are useful:
\begin{itemize}
\item They have enough power to pump optical parametric amplifiers (OPAs), to generate continua, or to drive self phase modulation broadening processes.
\item Operating at 100~kHz or higher drastically improves signal-to-noise and/or decreases the amount of time necessary for data collection, compared to previous generations of femtosecond lasers at 1-10 kHz.
\item High repetition rates enable efficient acquisition methods for data collection spanning femtoseconds to seconds.
\item High-repetition rates enable many ultrafast spectroscopies to be reengineered into hyperspectral microscopies.
\end{itemize}
In this article, we review the technology and present some recent applications, focusing on experiments performed by spectrometers operating at 100~kHz repetition rates.  At this repetition rate, Yb lasers have high pulse intensities, pulse shapers can alter individual laser pulses, and detectors can digitize signals shot-to-shot.

\section*{Overview of Y\lowercase{b}-based Amplified Laser Systems}

Yb laser systems, with outputs around 1030~nm, are poised to become the workhorse scientific short-pulse laser systems due to their advantages of high stability, high average power, and cost effectiveness. Owing to longer-wavelength emission, they can be pumped with high-power laser diodes directly, avoiding the additional diode-pumped frequency-doubled Nd:YAG lasers needed to pump Ti:Sa lasers. Yb’s low quantum defect of 0.09 when pumped at 940~nm, which determines how much of the pump energy is converted to heat, is half that of Nd (0.24 at 808~nm) and a third of Ti:Sa (0.33 at 532~nm), making it the ideal candidate for high average power diode-pumped solid-state lasers.\cite{Krupke2000}  The low quantum defect leads to low cooling requirements, improving robustness and removing complex cryo-cooling from equivalent power Ti:Sa amplifiers.
For the equivalent average output powers, a Yb system can easily use an order of magnitude less electrical power than a Ti:Sa system. This also comes with improved output quality - the significantly longer pulse energy correlation time in Yb systems compared to Ti:Sa, due to Yb’s longer fluorescence lifetimes (1~ms compared to 3.2~$\mu$s in Ti:Sa) has additional benefits on power stability. These performance improvements at high average powers have led to wide adoption by science and industry, with applications in micromachining and soft X-ray laser pumping further driving Yb laser developments.

Commercial high average power Yb chirped pulse amplification laser systems are now available using slab, thin disk, fibre or single crystal architectures. These provide watt- to kilowatt-level outputs\cite{Kramer2020} at 100~kHz with typical pulse lengths ranging from 300~fs to 1.2~ps.  While these systems cannot compete directly with Ti:Sa bandwidths, amplified Yb laser systems can generate comparable shortened pulses through self-phase modulation methods in hollow core fibres, multipass gas cells, and thin plates, as well as supercontinuum generation.\cite{Nagy2021} Direct conversion to the visible, via supercontinuum, can provide sufficient energy (tens of picojoules) for nonlinear spectroscopies,\cite{Kearns2017} while OPAs allow for higher pulse energies.

The broadened spectrum from nonlinear processes can be used to seed and pump OPAs, generating tunable wavelengths from the visible to the mid-IR. Importantly, OPAs use transparent crystals, making them resilient to the increases in average power at higher repetition rates. A variety of OPA designs have been used with Yb laser systems, covering many of the wavelengths of Ti:Sa pumped OPAs. To match the broad-bandwidths of Ti:Sa pumped OPAs, solutions have been demonstrated using OPCPA and noncollinear OPA (NOPA) approaches, increasing the amplified spectral bandwidth, and compressing using chirped mirrors, bulk materials, pulse shapers, and gratings.\cite{Liebel:14,Puppin:15,Luther2016,Thire:17,Budriunas:22}
The slightly higher fundamental wavelength of Yb (1030~nm) compared to Ti:Sa (800~nm) also improves the options of direct conversion to mid-IR wavelengths beyond 5~µm from the fundamental, using relatively new crystal materials.\cite{Penwell2018}

\begin{figure*}[t]
\includegraphics[width=0.8\textwidth]{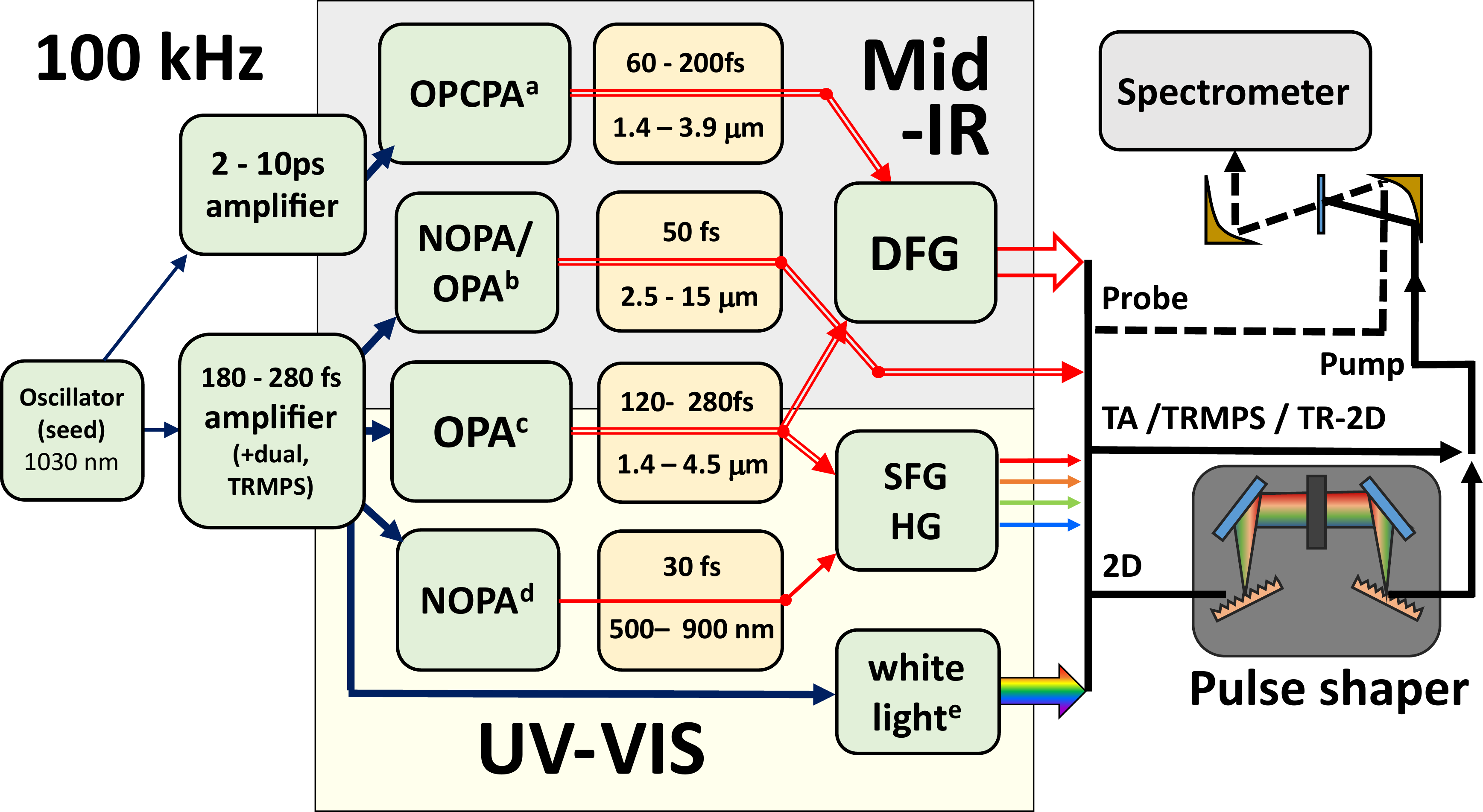}
\caption{There are many routes to femtosecond infrared and visible light sources operating at 100 kHz with pulse energies suitable for time-resolved spectroscopy. The conversion of the 1030 nm amplified light to mid-IR or visible can be via (a) OPCPA, (b) NOPA driven OPA, (c) OPA (d) NOPA  and (e) supercontinuum generation. The different operational stages are distinguished by arrow type. From left to right, we have weak 1030 nm seed light (thin navy arrow), amplified 1030 nm light (thick navy arrow), signal and/or idler (double/single red line), and the different visible and IR outputs as general sources of pump and probe for the kinds of spectroscopy mentioned in the text (black lines). DFG and SFG =difference/sum frequency generation, HG = harmonic generation.}  \label{figLasers}
\end{figure*}

Yb laser systems provide a flexible laser source for ultrafast spectroscopy measurements. This is illustrated in Fig.~\ref{figLasers}, which shows an overview of possible spectrometers, starting from a Yb oscillator as a seed. Downstream of the oscillator one can choose amplifier and conversion modules as needed for the spectroscopy technique of interest. We broadly divide the light sources shown here into  (a) OPCPA,\cite{Luther2016,Rigaud:16} (b) NOPA/OPA,\cite{Budriunas:22} (c) OPA \cite{Greetham2016, Donaldson2018, Farrell2020, Hamm2021}, (d) NOPA\cite{Liebel:14}  and (e) supercontinuum generation.\cite{Kearns2017}   For setting up time-resolved infrared spectroscopy measurements, one may choose to use the components noted in the top panel of Fig.~\ref{figLasers}, or  those in bottom panel, if one is interested in time-resolved electronic spectroscopy measurements.

\section*{Technology: Pulse Modulation and Detection Schemes}

Increased repetition rates necessitate improvements in pulse modulation and detection.  Intensity difference measurements, via pulse train modulation, is commonly used in ultrafast experiments to isolate small transient signals from large background signals.
Modulation schemes most commonly involve amplitude modulation, phase modulation, or a combination of the two.  Noise sources like laser noise, AC line noise or acoustic noise decrease with frequency.\cite{Anderson2007}  Therefore, it is generally true that modulating the signal at higher frequencies provides better S/N, with the limit being half the repetition rate of the laser.  The most common form of amplitude modulation is chopping with a mechanically rotating wheel comprising alternating open and closed slots.\cite{Kanal2014} 100~kHz mechanical chopping systems require small slots and tight focusing of light at the chopper. Problems of phase jitter and acoustic noise, a result of operation in air, can be eliminated by operation in vacuuum.\cite{Donaldson2016}

Optical phase modulation in ultrafast measurements is important for multidimensional spectroscopy.\cite{Myers:08, shim2009,Tan2008} To that end, most 100~kHz experiments to-date have used acousto-optic pulse shapers, which are a unique solution to optical modulation at 100~kHz repetition rates due to their ability to carry out amplitude and/or  phase modulation on a shot-by-shot basis.\cite{Middleton2010} In addition, pulse shapers can offset the material dispersion they introduce,\cite{Jones2019} and can rapidly scan over experimental variables such as pulse frequencies or pulse delays. Combining high-frequency phase modulation with rapid delay scanning has been shown to give higher S/N than other modulation schemes in pump-probe geometry 2D spectroscopy.\cite{Kearns2017}

In order to take full advantage of the S/N benefits that come from shot-to-shot modulation, it is also necessary to acquire the generated signals on a shot-to-shot basis.
Just like for lower-repetition rate instruments, it is highly advantageous to use array-detector systems, where probe light can be dispersed by a spectrometer onto the array, so that the full spectrum can be measured at once. Commercial array detector systems exist for 100~kHz UV to IR spectrum detection. For electronic spectroscopies, silicon line-scan CMOS linear arrays can provide $>$100~kHz data acquisition rates with thousands of pixels for UV to NIR.\cite{Kearns2017,Kanal2014}  In the mid-IR, liquid nitrogen cooled mercury cadmium telluride (MCT) arrays are commonly used with up to 128 pixels and can be used up to 100~kHz.\cite{Greetham2016,Luther2016,Hithell2016,Donaldson2018,Farrell2020}
The data rate at 100~kHz, even for large 1000 pixel array detectors, is not limiting with today's computers, however the slow electronic response time of the more common photoconductive MCT detectors (1-2~$\mu$s exponential decay time) limits the maximum repetition rate to about 100~kHz. To minimize crosstalk from one laser pulse to the next the electronic signal level should be below one percent of the total response which requires nearly 10~$\mu$s.

\begin{figure*}[t]
\includegraphics[width=0.8\textwidth]{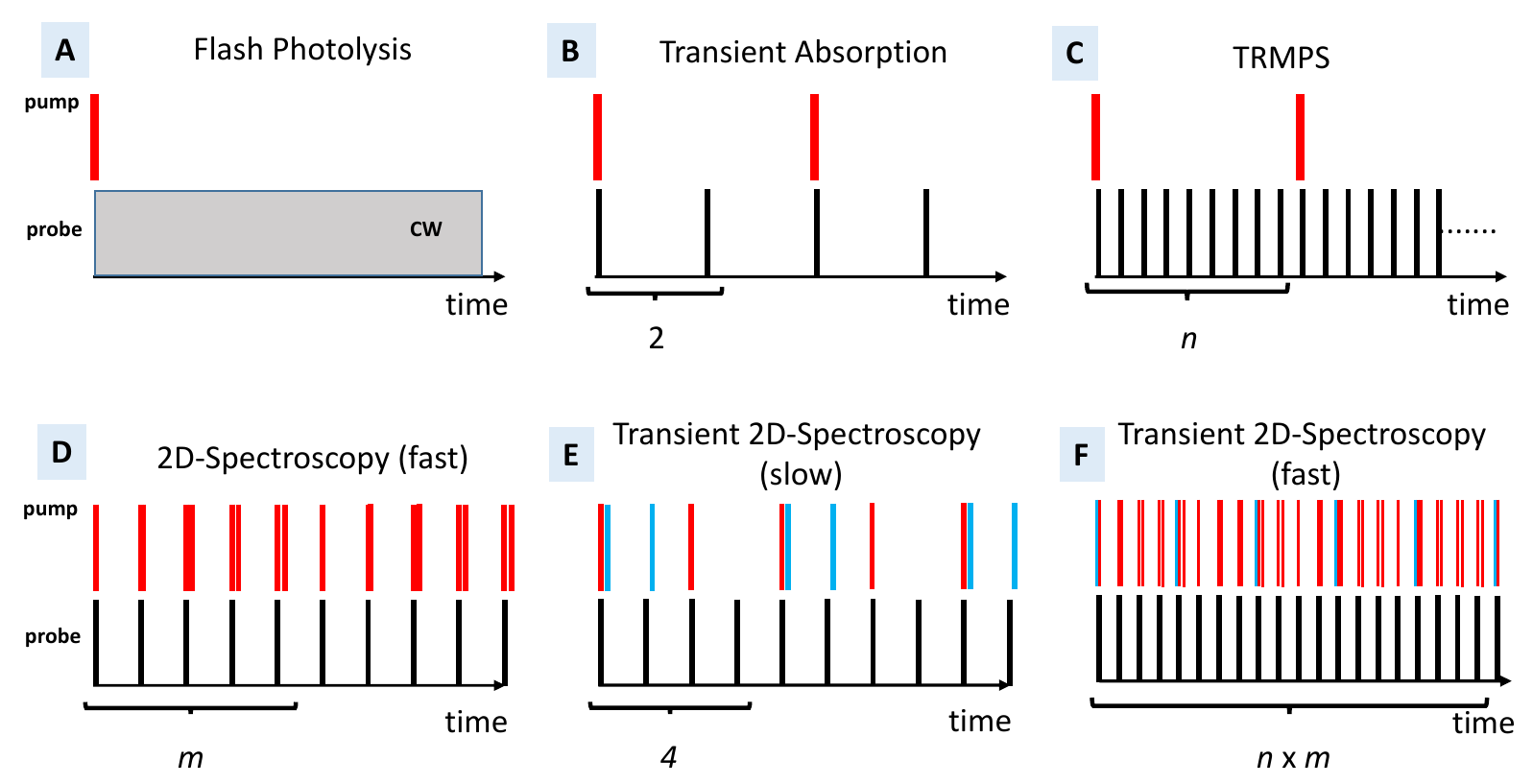}
\caption{Pulse schemes for time-resolved spectroscopy, as described in the main text. Each vertical line indicates a laser pulse. In each figure A-F, the pumping (top, red) and synchronous probing pulse configurations (bottom, black) are shown as a function of time.  In A, the grey box indicates continuous wave (CW) probing. In B-F, every probe pulse is measured. In transient 2D techniques E and F the actinic pump pulses are shown in blue. In all cases, averaged signals are acquired through repeated measurements of these diagrams.} \label{figPulseSequence}
\end{figure*}

\section*{100 \MakeLowercase{k}H\MakeLowercase{z} Lasers Open New Approaches to Time Resolved Spectroscopy}

Fig.~\ref{figPulseSequence} shows the principles of some of the time resolved spectroscopy techniques in-use. The figure describes processes where transient changes in sample absorption are probed. Variations not depicted could include Raman and surface sum-frequency generation for probing. For Fig.~\ref{figPulseSequence}, in all cases, probe light transmission changes tend to be in the range of 10$^{-3}$-10$^{-6}$ so regardless of the application, averaging over multiple repeated measurements is necessary. A higher repetition rate reduces measurement time or increases the signal-to-noise ratio of the measurement. However, the boost in repetition rate of Yb laser systems is so large that conceptually new experiments become feasible. We will focus our discussion below on some of those examples.

\subsection*{Time Resolved Multiple Probe Spectroscopy}

Many photochemical processes are initiated by the electronic excitation of a photoactive molecule, which then reacts extremely quickly on a femto- to picosecond timescale. This initial process is followed by a cascade of events that cover timescales up to seconds or even longer.
While the ultrafast photophysics of the initially excited compound in these molecular systems is typically understood from ``conventional'' femtosecond pump-probe experiments, studies of complete reaction cycles are scarce, since so far, they require the combination of different spectrometers.

Flash photolysis   gains time resolution by continuously measuring probe light intensity changes on a detector in ``real-time'' following optical pumping (Fig.~\ref{figPulseSequence}A). In the pump-probe technique, or ``transient absorption'' (TA, Fig.~\ref{figPulseSequence}B), temporal resolution is gained via exact control of the time delay $T$ between pulsed excitation and probe light. The short laser pulses define the time resolution, circumventing the detector-response limit,  resulting in a superior time resolution compared to flash photolysis.
Typically, two probe pulses are used per pump pulse, optically chopping the latter, in order to introduce a background measurement.

\begin{figure*}[t]
\includegraphics[width=0.9\textwidth]{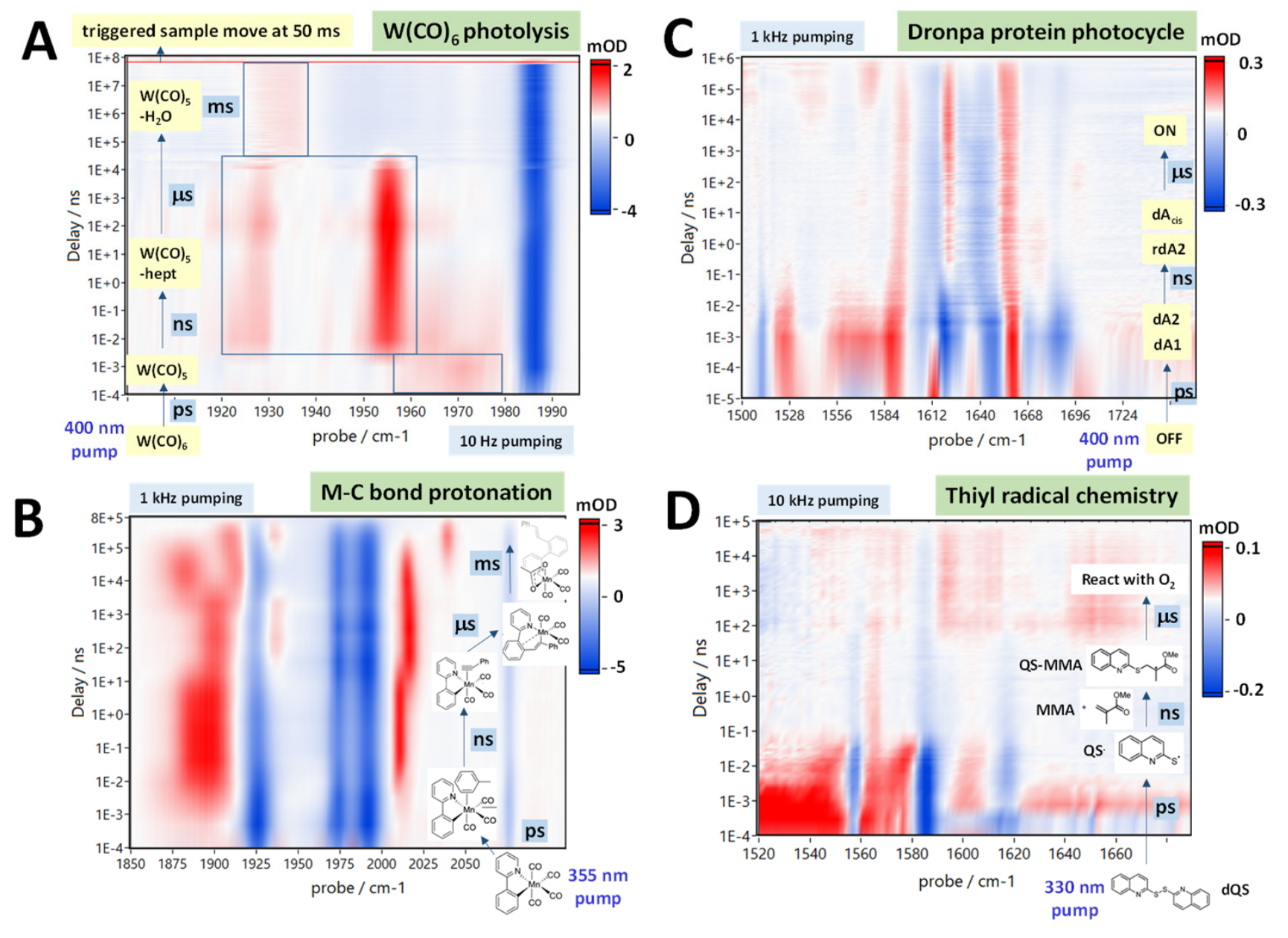}
\caption{100 kHz probe laser systems allow IR TRMPS across the widest time ranges, as illustrated in these examples: (A) The photolysis of W(CO)$_6$ is measured from picoseconds to 50 ms in a static solution of heptane. (B) Pico- to microsecond study of metal-carbon bond formation through the photolysis of Mn(ppy)(CO)$_4$ in toluene with dilute phenyl-acetylene and acetic acid. (C) Pico- to microsecond study of the Dronpa protein photocycle after pumping with 400 nm light. (D) Pico- to microsecond study of an example of thiyl radical chemistry.  All data were recorded using the CLF LIFEtime spectrometer. Data of (B) - (D) are from Refs.~\onlinecite{Hammarback2021,Laptenok2018,Koyama2017} with permission of the authors.   } \label{figTransientIT}
\end{figure*}

When a sample can withstand optical pumping at repetition rates of 100~kHz, and, after excitation, is fully relaxed in time for the next pump pulse, new applications of transient absorption spectroscopy to weak-signal samples become feasible, such as a recent time-resolved IR study of porphyrins in fixed HeLa tumour cells.\cite{Keane2022}  As many chemical samples of interest cannot be refreshed fast enough to withstand such a rate of pumping, a major potential of 100~kHz lasers we see with regard to TA spectroscopy lies in time resolved multiple probe spectroscopy (TRMPS, see Fig.~\ref{figPulseSequence}C).\cite{Greetham2012,Greetham2016,Jankovic2021b,Ruf2023} TRMPS combines the ``real-time'' measurement principles of flash photolysis with the TA benefit of ultrafast pulse probing, offering a far wider time-range in a single measurement than TA. The sampling of delay times $T$ spans fs-ns optical path-length delays, synchronised ns-$\mu$s pump laser delays and the delays of successive probe pulses separated by 10~$\mu$s each. TRMPS benefits from the exceptional long-term stability of 100 kHz Yb-based amplifier/OPAs and substantial signal-to-noise increases at long delay time.\cite{Greetham2016}

In Fig.~\ref{figTransientIT}A, the well-known example of W(CO)$_6$ photolysis followed by solvent adduct formation\cite{KELLY1974} is shown in a 100~kHz TRMPS measurement spanning a sub-picosecond to 50~ms time-frame. The sample was pumped at 400~nm and 10~Hz, and probed at 100 kHz in the mid-IR.\cite{Greetham2016} A synchronised, fast $xyz$-positioner was used to refresh the sample between pump pulses (stop-flow can also be used to refresh the sample in these type of experiments\cite{Buhrke2021}). Each spectrum at each unique time delay was an average of 50 pump shots, demonstrating an excellent signal-to-noise ratio. The bleach of the parent W(CO)$_6$ can be seen to persist indefinitely, and the sequential formation of the heptane (ps) and water adducts ($\approx$1~ms) are clearly visible.

Fig.~\ref{figTransientIT}B  shows the  reaction of Mn(ppy)(CO)$_4$ in toluene with dilute phenyl-acetylene and dilute acetic acid.\cite{Hammarback2021,Hammarback2022} After photoexcitation, the toluene solvent adduct forms, followed by replacement with phenyl-acetylene on a nanosecond timescale. Intra-molecular reaction with the 2-phenylpyridyl ligand takes place on the microsecond timescale and protonation of the nascent cyclometalated ligand via acetic acid occurs on the millisecond timescale, resulting in an acetate coordinating to the Mn.\cite{Hammarback2021}
Fig.~\ref{figTransientIT}C shows a 100~kHz TRMPS application to the study of the ps-$\mu$s steps of the Dronpa protein photocycle.\cite{Laptenok2018}  The transformation of the UV absorbing ‘OFF’ state of the photoactive Dronpa protein to its emissive, blue absorbing ‘ON’ state, captured by 100 kHz TRMPS-IR takes hundreds of $\mu$s. Finally, Fig.~\ref{figTransientIT}D shows the ultrafast generation and ns-$\mu$s reactions of quinolin sulphide (QS) radicals with solvent methyl methacrylate (MMA).\cite{Koyama2017}

Other examples of transient chemistry explored using the 100 kHz TRMPS-IR technique include peptide unbinding from RNase S,\cite{Jankovic2021b} the molecular mechanism of light-induced bond formation in a cyanobacteriochrome,\cite{Ruf2023} photo-catalytic decarboxylation reactions,\cite{Bhattacherjee2019} radical induced 1,2 metalate rearrangement reactions,\cite{Lewis-Borrell2021}   and photoinduced polymerisation reactions.\cite{Koyama2017}

\subsection*{2D Spectrometers Using 100 kHz Amplified Laser Systems}

Coherent two-dimensional spectroscopy measurements have transformed the quantitative details that can be extracted from molecular systems across a multitude of scientific fields ranging from biology to materials science and beyond.  In the past 25 years, 2D electronic spectroscopy (2D ES) and 2D~IR spectroscopy have been used to investigate light harvesting systems, protein dynamics and structures in complex environments, as well as ultrafast hydrogen-bonding dynamics in condensed-phase environments, to name a just a few applications.\cite{hamm11,Nuernberger2015,Ghosh2017,Collini2021} 2D spectroscopy in the pump-probe geometry is in essence a TA experiment with the dependence on pump frequency determined, requiring extra measurements of separate pump frequencies or interferograms (Fig.~\ref{figPulseSequence}D).\cite{shim2009}

\begin{figure*}[t]
\includegraphics[width=0.80\textwidth]{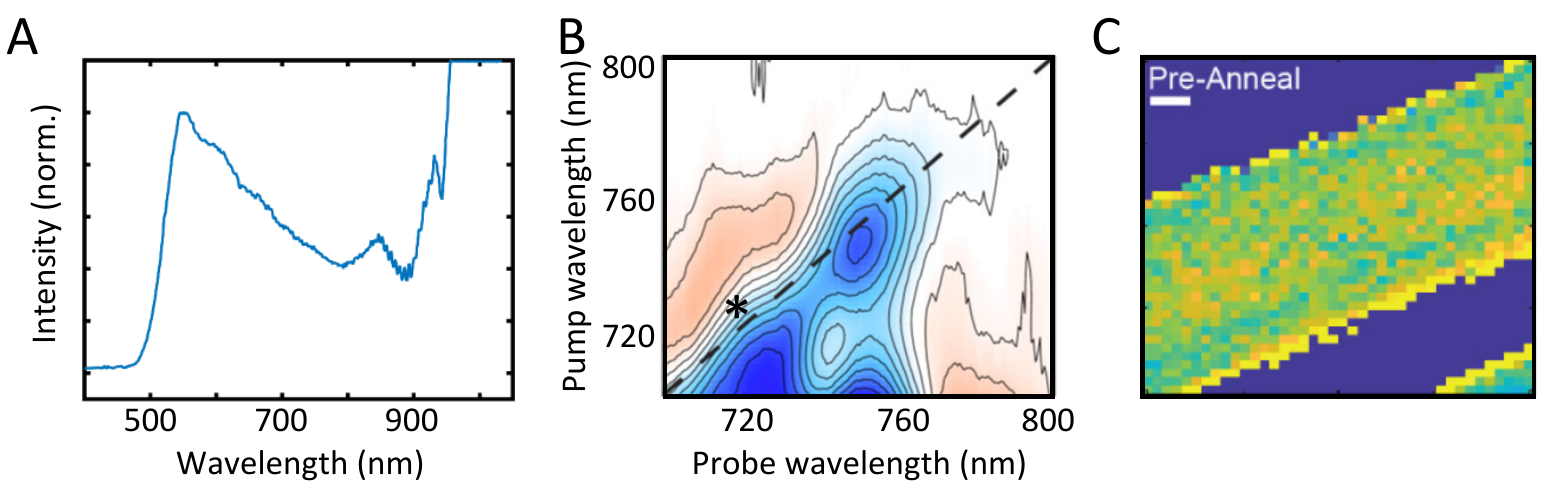}
\caption{(A) Example white-light continuum generated from focusing the Yb laser output into an 8~mm YAG crystal. (B) 100 kHz 2D WL spectrum of a methylammonium-PbI$_3$ perovskite thin film. (C) Broadband TA image of a TIPS-pentacene microcrystal that spatially maps slip-stacked structures. The colors represent the percentage of slip-stacked non-equilibrium structures with yellow being the maximum population of 16 percent. The scale bar is 2 $\mu$m. Adapted with permission from references~\onlinecite{Armstrong2020,Kunz2021}, Copyright 2020 and 2021 American Chemical Society.} \label{fig2DVis}
\end{figure*}

High-repetition rate lasers are well-suited for spectroscopies that use a continuum as the pump source.  For transient absorption spectroscopy, white-light continua have been used for decades as a probe light source, usually generated by focusing a small amount of the output of a Ti:Sa regenerative amplifier into a liquid, solid or gas.
Similar principles can be used to generate continua in other regions of the optical spectrum. For TA experiments, the prevailing idea was to use an intense pump pulse and a non-perturbative, i.e. weak, probe pulse. Thus, the pump pulse was usually generated by an optical parametric amplifier to create intense and tunable pulses that selectively excite a particular electronic state.

For 2D spectroscopy, in contrast, one wants as wide as possible spectrum in both dimensions, i.e. a continuum also for the pump pulse. As the bandwidth  is increased in both dimensions, more off-diagonal features can be observed.  Continua have been used in 2D~IR experiments with low repetition-rate lasers,\cite{Hack23, Ramasesha2013}  and in 2D White-Light (2D WL) spectroscopy experiments performed with high repetition-rate lasers.\cite{Armstrong2020} Shown in Fig.~\ref{fig2DVis}A is the spectrum of white-light generated using about 2~$\mu$J of $\approx$200~fs light from a 100~kHz Yb laser. The white-light turns on at about 540~nm and spans to about 1300~nm, a bandwidth far broader than that of any NOPA. Of course, the white-light generated from only a few $\mu$J's is itself $<$1~µJ.  Therefore, it may not be suitable for all applications, but many systems of interest in biology, chemistry and material science are strong absorbers: the purpose of leaves and photovoltaics is to absorb light, after all. In addition, the exceptional noise performance of 100~kHz Yb lasers compensates for the low pulse energy.

Shown in Fig.~\ref{fig2DVis}B is an example of a 2D WL spectrum collected with such a continuum pump. It is of a lead halide pervoskite thin film, resolving a previously unreported feature 150 meV above the bandgap (marked as * in Fig.~\ref{fig2DVis}B)  that does not appear in analogous Br$^-$/Cs$^+$ films, suggesting that the electronic structure is altered to impact efficiency.
We note that this experiment does not utilize any OPA; white light generated from a 100~kHz laser alleviates the need for an OPA.

\begin{figure*}[t]
\includegraphics[width=.9\textwidth]{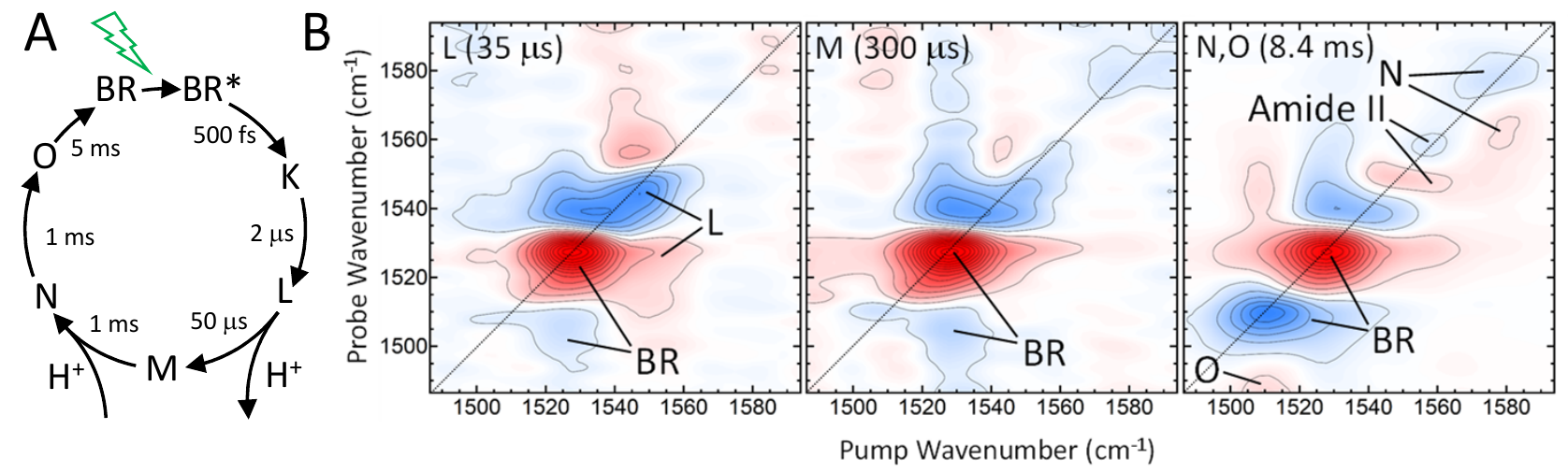}
\caption{(a) Photocycle of bacteriorhodopin. After optical excitation, \textit{trans-cis} isomerisation of one of the C=C bonds of the retinal chromophore results in the K-state, followed by a sequence of intermediates (L, M, N and O) which include de-protonation and re-protonation of the retinal Schiff-base, making it a proton-pump. (b) 100 kHz transient 2D~IR spectra of the photocycle of bacteriorhodopsin taken at time points when the populations of the various intermediates peak. Adapted with permission from Reference~\onlinecite{Hamm2021}, Copyright 2021 American Institue of Physics.} \label{figTransient2DIR}
\end{figure*}

Another example of a 2D experiment, only possible with a 100~kHz laser, is shown in Fig.~\ref{figTransient2DIR}. In this case, the very ideas of high-repetition rate TRMPS and 2D~IR spectroscopy have been combined for a new form of transient 2D~IR spectroscopy, covering timescales from 10~$\mu$s to 10's of milliseconds.\cite{Hamm2021,Buhrke2023}  In ``conventional''  transient 2D~IR spectroscopy (Fig.~\ref{figPulseSequence}E), three beams hit the sample:\cite{bredenbeck2003} one beam containing an actinic UV/VIS pulse to excite a photochemical reaction in the sample, which is followed by a 2D~IR pulse sequence in the pump-probe geometry. To single out the desired  transient 2D~IR response, two choppers are included defining 4 measurement states with both the UV/VIS-pump either on or off and the same for the IR-pump pulses. For Fig.~\ref{figPulseSequence}E, the maximum time-range accessible with this approach is ultimately limited by the repetition rate of the laser system, and a high repetition rate is not necessarily advantageous. Furthermore, the 2D~IR pulse sequence is step-scanned, which is problematic in terms of signal-to-noise ratio and scatter suppression.

In Ref.~\onlinecite{Hamm2021}, an approach has been presented which intertwines the scanning of the actinic pump pulse in steps of 10~$\mu$s (i.e., derived from a 100~kHz laser) with that of the two IR pump-pulses for the 2D~IR pulse sequence; the latter with a fast-scanning pulse shaper. Both are scanned in an asynchronous manner (Fig.~\ref{figPulseSequence}F) in a way that each of the two delay times appear in the shortest possible measurement time (a few seconds). One can then extract a sequence of thousands of 2D~IR spectra in steps of 10~$\mu$s after the actinic pump pulse.

Bacteriorhodopsin is a proton-pump, that is driven by the light-induced \textit{trans-cis} isomerisation of its retinal chromophore. While the isomerisation occurs within only 500~fs, the complete photocycle to recover the original state of the chromophore takes about 5~ms (see Fig.~\ref{figTransient2DIR}a). Fig.~\ref{figTransient2DIR}b shows a sequence of  transient 2D~IR spectra taken at which the populations of the various intermediates during the photocycle peak.\cite{Hamm2021} The transient 2D~IR spectra show the evolution of the spectroscopic properties of the retinal chromophore and its coupling to the amide II band of the protein.

\subsection*{100 kHz Lasers Make 2D Imaging Experiments Practical}

\begin{figure*}[t]
\includegraphics[width=0.9\textwidth]{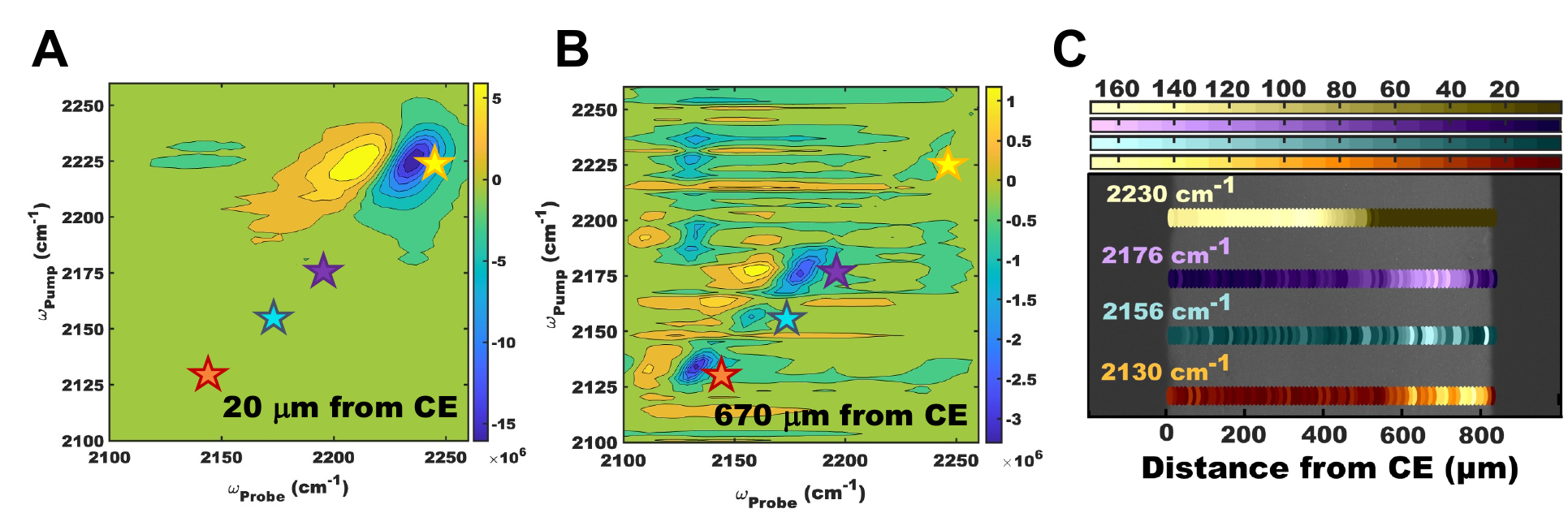}
\caption{(a-b) 2D~IR spectra extracted from 2D~IR imaging of a copper electrochemical cell.  The stars indicate the peaks monitored as a function of distance from the CE.  (c) Brightfield image taken of a section of the copper electrochemical cell.  The scaled and integrated intensity for the $\nu=0-1$ transitions of [DCA]$^-$ modes centered at 2230, 2176, 2156, and 2130 cm$^{-1}$ are overlaid in gold, purple, blue, and orange dots, respectively.  The $y$-positions  are offset for visualization only.   Adapted with permission from reference~\onlinecite{Tibbetts2021}, Copyright 2021 American Chemical Society   } \label{Fig6_2DIRImage_Final}
\end{figure*}

In non-homogeneous samples, resolving the spatial heterogeneity of samples is often necessary to fully understand the chemical dynamics. Nonlinear spectroscopies can be turned into microscopy experiments by tightly focusing the laser beams and raster scanning the sample, thus creating an image that is a collection of many independently measured spectra. Generating hyperspectral images at 1 kHz is possible for some samples, such as animal tissues, but extremely time-consuming.\cite{Baiz2014,Ostrander2016,Alperstein2021,Dicke2021}  The limiting factors are signal-to-noise and the time it takes to collect each spectrum, which are interdependent. In the opposite extreme, ultrafast TA microscopy experiments have used lasers with MHz repetition rates, but modulated at 10's of kHz together with lock-in detection that effectively reduces the repetition rate, and is incompatible with array detectors.
100 kHz 2D microscopes were first implemented in the infrared and later in the visible utilizing broadband probes, pulse shapers and shot-to-shot detection.\cite{Tibbetts2021,Jones2019b} Both used point-scanning and a fully collinear optical geometry.

Shown in Fig.~\ref{Fig6_2DIRImage_Final} is an example of 2D~IR microscopy used to spatially resolve copper complexes in an electrochemical cell made of copper electrodes and a room-temperature ionic liquid (i.e.,1-butyl-3-methylimidazolium tetrafluoroborate and 1-butyl-3-methylimidazolium dicyanamide, [BmimBF4] and [BmimDCA], respectively) electrolyte system. A potential of -2.5~V with respect to the copper counter electrode (CE) was held constant through the course of the 2D~IR imaging experiment.  2D~IR spectra of the copper [DCA] complexes evolve as the focus of the laser is scanned from the CE toward the working electrode (WE) (see Fig.~\ref{Fig6_2DIRImage_Final}).  Analysis of these 2D~IR spectra show that there is a variation in chemical species detected as a function of distance from the electrode.\cite{Tibbetts2021} It is possible that these variations are caused by ultrafast chemical exchange whose dynamics depend on potential-dependent concentrations across microscopic length scales and much longer time scales as the coordinated copper complexes are transported from the CE to the WE. These types of experiments have the potential to unveil how potential dependent chemical dynamics drive important battery chemistry including chemistry associated with the formation of solid electrolyte interphases.

In an instrument analogous to the 2D~IR microscope described above, but using a 2D WL spectrometer, images have been measured of singlet-fission microcrystals, resolving previously unseen slip-stacked structures (see Fig.~\ref{fig2DVis}c).\cite{Armstrong2020,Jones2020}
And in work underway, spatial variations in 2D perovskites are being measured with 2D WL imaging.

\section*{Outlook}

The experiments using 100kHz Yb amplifiers discussed here have greatly impacted our laboratories.  Their performance has enabled experiments not previously possible, and their reliability has improved our productivity.  These two aspects have been especially felt at the Rutherford Appleton Laboratory's Central Laser Facility, a  user facility in the UK where Yb lasers have expanded the facility's  focus on femtosecond dynamics to measurements out to seconds , thereby better serving a large range of scientific disciplines.

This review focused on experiments performed at 100 kHz repetition rate, where Yb lasers can pump OPAs or generate white-light, where pulse shapers exist that can create a different pulse train with each laser shot, and linear array detectors can be read out shot-to-shot.   Yb lasers are improving rapidly, as are changes to these other pieces of equipment, and so repetition rates, energies and bandwidths will be pushed higher as well.  Of course, there is no one spectrometer design for all experiments, but we find it interesting that the breadth of experiments contained in this review are all accomplished with similar experimental apparatus design.

Besides the examples discussed here, there are many other ideas for how non-linear spectroscopy that is enabled by Yb technology might impact the larger scientific community. There are many applications of 2D~IR spectroscopy where the better signal-to-noise ratio provided by 100~kHz systems allows one to tackle samples that otherwise would be nearly impossible.
For the same reason, many other ultrafast experiments  should be made significantly more practical at higher repetition rates. Already, 100~kHz laser systems have been used to decrease acquisition times for 2D-Raman-THz spectra from days to hours,\cite{Duchi2021} and for 2D-IR-Raman spectra, a novel Raman analogue of conventional 2D~IR spectroscopy, from hours to minutes.\cite{Donaldson2020}

Yb technology could transform experiments that require scanning two time delays , such as action-detected 2D spectroscopies. In these experiments, a final ``action'' measurement is performed after a sequence of 4 pulses that include the two coherence times of a 2D experiment. The action measurement can be almost any incoherent process, such as fluorescence,\cite{Tekavec2007,Draeger:17},
photocurrent,\cite{Nardin:13,Karki2014}
mass spectrometry,\cite{Roeding2018,Chen2021} or photoelectron  detection.\cite{Aeschlimann2022}

Another example in this regard is 3D spectroscopy requiring three coherence times scanned, only one of which can be taken care of by array detection.\cite{Ding2007,Garrett-Roe2009,Turner2009,Zhang:12}
Also, heterodyne-detected 2D SFG spectroscopy, which has not seen wide-spread adoption like its 1D counterpart due to the very small signal sizes.\cite{Xiong2011,Singh2012,Wang2015}
Faster acquisition times should also enable rapid-scan 2D and TA spectroscopy examining irreversible processes such as mixing on-chip,\cite{Tracy2016} fibril formation,\cite{Buchanan2013} polymerization, and crystallization. Rapid screening applications are now within reach as well.\cite{Fritzsch2018}

In summary, we are convinced that Yb-based 100~kHz laser systems will soon become a standard of femtosecond laser technology, replacing many Ti:Sa lasers applications. Their improved  parameters will not only make already existing experiments easier, better, and faster, but will also allow for conceptually new experiments. As well as breaking barriers to new spectroscopies, Yb lasers also lower barriers to starting an ultrafast spectroscopy laboratory. While we have presented here some examples, only the future will fully unfold the potential of this new technology.\\

\noindent\textbf{Acknowledgement}  ATK acknowledges support from the Air Force Office of Scientific Research (FA9550-20-1-0401) and the Department of Energy (DE-SC0016137), MTZ from the Air Force Office of Scientific Research (FA9550-19-1-0093) and the National Science Foundation (CHE 1665110), PH from the Swiss National Science Foundation (NCCR MUST and 200021\_214809), GMG and PMD from STFC and BBSRC for the LIFEtime instrument (BB/L014335/1), and PMD from UKRI (MR/S015574/1).  We thank all our coworkers who contributed to the research.\\

\noindent\textbf{Conflict of Interest} MTZ and CTM are co-owners of PhaseTech Spectroscopy, Inc., which manufactures instruments similar to those used here.\\

\noindent\textbf{Biographies}

Paul M. Donaldson (PhD 2007, Imperial College London) is a UKRI future leader fellow at the UK Central Laser Facility.

Greg M. Greetham (Ph.D. 2000, University of Leicester) is a Group Leader at the UK Central Laser Facility.

Chris T. Middleton (Ph.D. 2008, Ohio State University) did his postdoctoral work at the University of Wisconsin-Madison. He is co-founder, and currently Vice-President and CTO of PhaseTech Spectroscopy, Inc.

Bradley M. Luther (Ph.D. 2000, Colorado State University) is a Research Scientist at Colorado State University.

Martin T. Zanni (Ph.D. 1999, University of California, Berkeley) is the Meloche-Bascom Professor at the University of Wisconsin-Madison and co-founder of PhaseTech Spectroscopy, Inc.

Peter Hamm (Ph.D. 1995, Ludwig Maximilians University Munich) is Professor at the University of Zurich since 2001.

Amber Krummel (Ph.D. 2007, University of Wisconsin—Madison) is Professor at Colorado State University since 2010.



%

\end{document}